\documentclass[crcready]{iosart2x}

\usepackage{dcolumn}
 \usepackage{lscape}
 \usepackage{longtable}
 \usepackage{graphicx}

\pubyear{0000}
\volume{0}
\firstpage{1}
\lastpage{1}

\begin{document}
\begin{frontmatter} 
\onecolumn
%
\title{Exploring ChatGPT for Next-generation Information Retrieval: Opportunities and Challenges}


\begin{aug}
\author{\inits{Y.}\fnms{Yizheng} \snm{Huang}\ead[label=e1]{hyz@yorku.ca}}
\author{\inits{J.}\fnms{Jimmy X.} \snm{Huang}\ead[label=e2]{jhuang@yorku.ca}%
\thanks{Corresponding author. \printead{e2}.}}
\address{Information Retrieval and Knowledge Management Research Lab, \orgname{York University},
Toronto, \cny{Canada}\printead[presep={\\}]{e1,e2}}
\end{aug}

\begin{abstract}
The rapid advancement of artificial intelligence (AI) has highlighted ChatGPT as a pivotal technology in the field of information retrieval (IR). Distinguished from its predecessors, ChatGPT offers significant benefits that have attracted the attention of both the industry and academic communities. While some view ChatGPT as a groundbreaking innovation, others attribute its success to the effective integration of product development and market strategies. The emergence of ChatGPT, alongside GPT-4, marks a new phase in Generative AI, generating content that is distinct from training examples and exceeding the capabilities of the prior GPT-3 model by OpenAI. Unlike the traditional supervised learning approach in IR tasks, ChatGPT challenges existing paradigms, bringing forth new challenges and opportunities regarding text quality assurance, model bias, and efficiency. This paper seeks to examine the impact of ChatGPT on IR tasks and offer insights into its potential future developments.
\end{abstract}

\end{frontmatter}

\section{Introduction}\label{intro}
On November 30, 2022, OpenAI unveiled ChatGPT\footnote{https://chat.openai.com}, an AI chatbot application powered by the advanced GPT-3.5 and later GPT-4 generative language models. This application quickly attracted over a hundred million users worldwide, setting a new record for rapid product dissemination \cite{openai2023gpt4}. ChatGPT, as an embodiment of these models, demonstrated significant advancements over its predecessors, quickly becoming a central topic in both industrial and academic circles. While some view ChatGPT as a disruptive technological innovation, predicting revolutionary changes in various sectors, others believe its success stems more from effective product and market strategies than from purely technological breakthroughs.

Indeed, ChatGPT heralded a new phase in Generative AI, distinct from previous models like GPT-3. This new generation of AI models, including ChatGPT, is capable of generating unique content, not just refining or predicting information based on training examples \cite{deng2022benefits}. GPT-3.5 established a strong foundation with its robust capabilities \cite{schulman2017proximal, brown2020language, neelakantan2022text}. GPT-4 further expanded these capabilities, offering enhanced understanding, accuracy, and contextual relevance. The evolution from GPT-3.5 to GPT-4 has shown great promise in numerous information retrieval tasks (e.g. \cite{DBLP:journals/ipm/HuangMH13, DBLP:journals/jasis/YeHL11}), particularly in text classification \cite{DBLP:conf/icdm/HuangHWALP06}, document ranking \cite{DBLP:conf/trec/HuangZS05}, question-answering systems \cite{DBLP:conf/sigir/ChenHHHA17}, and multimodal retrieval \cite{10.1145/3477495.3531722}. The introduction of ChatGPT, leveraging these advancements, has spurred progress in this field, highlighting the impressive abilities of large language models (LLMs) in understanding and generating semantic information.

Amid these rapid technological developments, ChatGPT has been applied in various practical settings. Notably, it powers Microsoft's AI-driven search engine, New Bing, based on GPT-4\footnote{https://blogs.microsoft.com/blog/2023/02/07/reinventing-search-with-a-new-ai-powered-microsoft-bing-and-edge-your-copilot-for-the-web}, and integrates with other multimodal pre-trained models, enhancing the scope of IR tasks. Traditionally, supervised learning has been the main approach in IR, involving the design of statistical or probabilistic models trained on specific task-related data, parameter optimization through loss function minimization, and model inference on new data. The advent of deep neural networks shifted the focus from traditional machine learning models to deep learning models. However, the reliance on the supervised learning framework persisted. This method, training models on labeled datasets to predict or categorize unseen data, has driven significant progress in various IR applications. Nonetheless, the emergence of ChatGPT and the GPT-X models it is based on (where X represents different versions) has posed new challenges to existing IR paradigms, introducing research and application issues such as ensuring text quality, addressing model bias and ethical concerns, and improving model efficiency and practicality.

This paper delves into the opportunities and challenges brought forth by ChatGPT in IR tasks. We also offer a forward-looking view on the future development of ChatGPT and its underlying GPT-X models, aiming to provide valuable insights for research and applications in related fields.

\begin{table*}[t]
\centering
\caption{Comparison of pre-trained large language models in recent years.}
\label{tab:compar_pllms}
\begin{tabular}{|c|c|c|c|}
\hline
\textbf{Pre-trained Language Models} & \textbf{Release Data} & \textbf{Size of Pre-training Corpus} & \textbf{Parameters Size} \\ \hline
BERT-Large \cite{DBLP:conf/naacl/DevlinCLT19}  & 2018-10 & 16 GB            & 340M  \\ \hline
GPT-2 \cite{radford2019language}      & 2019-02 & 40GB             & 1.5B  \\ \hline
RoBERTa  \cite{liu2019roberta}   & 2019-07 & 161 GB           & 340M  \\ \hline
XLNet-Large \cite{yang2019xlnet} & 2019-07 & 142 GB           & 340M  \\ \hline
T5-11B \cite{raffel2023exploring} & 2019-10 & 750 GB           & 11B  \\ \hline
OPT \cite{zhang2022opt}         & 2020-05 & 180B tokens      & 175B  \\ \hline
GPT-3 \cite{brown2020language}       & 2020-06 & 45TB             & 175B  \\ \hline
mT5-XXL \cite{xue2021mt5}     & 2020-10 & 750 GB           & 13B   \\ \hline
ERNIE 3.0 \cite{sun2021ernie}   & 2021-07 & 375B tokens      & 10B   \\ \hline
Yuan 1.0 \cite{wu2021yuan}    & 2021-10 & 180B tokens      & 245B  \\ \hline
PaLM \cite{chowdhery2022palm}       & 2022-04 & 780B tokens      & 540B  \\ \hline
BLOOM \cite{workshop2022bloom}      & 2022-11 & 366B tokens      & 176B  \\ \hline
GPT-4 \cite{openai2023gpt4}      & 2023-04 & About 13T tokens & About 1.76T \\ \hline
PaLM2 \cite{chowdhery2022palm}       & 2023-05 & 100B tokens      & 16B   \\ \hline
LlaMA2 \cite{Touvron2023Llama2O}      & 2023-07 & 2T tokens        & 70B   \\ \hline
Qwen-14B \cite{bai2023qwen}      & 2023-09 & 2.4T tokens        & 14B   \\ \hline
Skywork \cite{wei2023skywork}     & 2023-10 & 3.2T tokens      & 13B   \\ \hline
\end{tabular}
\end{table*}

\section{Pretrained Large Language Models}

The field of information retrieval has experienced a remarkable transformation with the emergence of pretrained large language models (PLLMs). This evolution, progressing from initial simplistic models to the current advanced dense retrieval models, has significantly broadened the scope and capabilities of IR and related fields. A comparison of recent pre-trained language models, including their training datasets and parameter size, can be seen in Table \ref{tab:compar_pllms}.

\paragraph{Early Language Models} The era of language models began with statistical approaches, notably n-gram models. These models predicted subsequent words based on the probability distribution of word sequences in sentences. The field advanced with the introduction of neural network-based models, such as the Neural Probabilistic Language Model \cite{bengio2000neural}, marking a new phase in language modeling. Following this, architectures like Convolutional Neural Networks (CNNs) \cite{gu2018recent}, Recurrent Neural Networks (RNNs), and Long Short-Term Memory Networks (LSTMs) \cite{sherstinsky2018fundamentals} emerged. These architectures addressed issues like data sparsity and capturing long-term dependencies but faced challenges in processing long sequences and parallelization.

\paragraph{The Transformer Paradigm} A significant breakthrough occurred with the introduction of the Transformer architecture by Google in 2017 \cite{vaswani2017attention}. This architecture, featuring self-attention mechanisms, enabled efficient parallel processing of sequences and effective management of long-term dependencies, overcoming many limitations of previous models.

The evolution of OpenAI's Generative Pre-trained Transformer (GPT) series is a testament to the success of the Transformer architecture. GPT-1 laid the foundation, and subsequent versions, GPT-2 and GPT-3, dramatically expanded the scale and capabilities of these models. Notably, GPT-3, with its 175 billion parameters, demonstrated an impressive leap in generating human-like text and facilitating meaningful interactions.

\paragraph{ChatGPT: The New Frontiers} Building on GPT-3, OpenAI developed ChatGPT based on the GPT-3.5 architecture. This model was specifically designed to overcome certain limitations of GPT-3, particularly in producing coherent and contextually relevant responses over extended dialogues. The training of ChatGPT involved a novel approach, Reinforcement Learning from Human Feedback (RLHF) \cite{ouyang2022training}, involving multiple iterations of model refinement using a reward model created from human-ranked responses.

ChatGPT represents a significant advancement in creating models capable of more meaningful and context-aware user interactions. Its deployment has demonstrated potential for a wide range of real-world applications, as highlighted by recent studies and deployments \cite{li2023chatdoctor, park2023generative, DBLP:conf/acl/LaskarBRBJH23, DBLP:conf/bionlp/JahanLPH23, laskar-etal-2023-large}.

\paragraph{Training Methodologies of ChatGPT} ChatGPT's architecture, based on the transformer model, includes specific modifications to enhance conversational abilities. The RLHF training method is notable, involving human trainers who guide the model by ranking responses, thereby refining the model's capability to generate contextually appropriate responses. This training also incorporates safety and bias reduction measures, ensuring adherence to ethical guidelines.

\paragraph{Interaction Mechanisms with Prompts} ChatGPT's interaction with prompts involves understanding user input's intent and context. It generates responses that are relevant, coherent, and contextually suitable by combining learned patterns from its training data with real-time input processing. This process also includes managing ambiguous or incomplete information and maintaining context over a conversation. 
\paragraph{GPT4: Advancing ChatGPT's Capabilities} GPT-4, released after ChatGPT and GPT-3.5, further pushes the boundaries of PLLMs. With an extended context window and hypothesized multimodal capabilities, GPT-4 is posited to surpass GPT-3.5 in many respects, potentially matching or exceeding human performance in various tasks. Its extendibility is evident in integrations and new services like Microsoft's Copilot\footnote{https://blogs.microsoft.com/blog/2023/03/16/introducing-microsoft-365-copilot-your-copilot-for-work}, enhancing productivity tools.

\paragraph{ChatGPT in IR} ChatGPT significantly contributes to IR by understanding and responding to queries using its extensive internal knowledge base. Unlike traditional search engines, ChatGPT simplifies the user experience by generating useful answers without requiring users to have specific knowledge, making it an invaluable tool for various tasks. A typical scenario is ChatGPT's robustness in understanding queries that contain grammatical or spelling errors. Even when a user submits a query with such inaccuracies, ChatGPT effectively interprets the intended meaning and provides responses incorporating correct grammar and spelling. This feature enhances user experience, ensuring that communication barriers due to language proficiency or typing errors do not hinder the retrieval of accurate and relevant information.

\paragraph{Other Noteworthy Models} The PLLM landscape features several key players besides the GPT series, as shown in Table \ref{tab:compare_chat}. ChatGPT is renowned for its creative text generation and remarkable scalability with plugins. Nevertheless, it has the problem of producing incoherent or incorrect text. Meta's LlaMA-2\footnote{https://ai.meta.com/llama} has a range of parameter sizes and offers versions that are fine-tuned for specific tasks. Despite the various parameters available, the model lacks a user-friendly bot interface, limiting access to the normal user. Google's Bard\footnote{https://bard.google.com} stands out for its ability to respond consistently to varied queries but with limited creativity. Lastly, Anthropic's Claude\footnote{https://www.anthropic.com/index/introducing-claude}, while not fully disclosing its architecture, has drawn attention for its extensive token capacity, facilitating the processing and generation of lengthy and complex texts. In addition, Claude is committed to reducing the generation of false or misleading information. However, it operates under a strict content review strategy, which may restrict access to legitimate information, particularly in fields like scientific research. These models reveal unique strengths and challenges, contributing to the dynamic PLLM field. 

\begin{table*}[t]
\centering
\caption{Comparison of ChatGPT, Llama-2, Bard, and Claude.}
\label{tab:compare_chat}
\resizebox{\textwidth}{!}{%
\begin{tabular}{|c|c|c|c|c|}
\hline
\textbf{Models} &
  \textbf{Company} &
  \textbf{Architecture} &
 \textbf{Notable Strengths}  &
  \textbf{Notable Weaknesses} \\ \hline
ChatGPT &
  OpenAI &
  Generative Pre-trained Transformer (GPT) &
  \begin{tabular}[c]{@{}c@{}}Creative text generation\\ Scalability (e.g. integration with DALL-E)\end{tabular} &
  Generate incoherent or incorrect text \\ \hline
Llama-2 &
  Meta &
  Auto-regressive Language Optimized Transformer &
  \begin{tabular}[c]{@{}c@{}}Range of parameter sizes (7B, 13B, and 70B)\\ Fine-Tuned Versions\end{tabular} &
  No convenient bot-like interface \\ \hline
Bard &
  Google &
  Pathways Language Models (PaLM2) \cite{chowdhery2022palm} &
  \begin{tabular}[c]{@{}c@{}}Faster\\ Coherent responses\end{tabular} &
  Significant creative limitations \\ \hline
Claude &
  Anthropic &
  Not fully disclosed &
  \begin{tabular}[c]{@{}c@{}}Large token capacity\\ Reduced hallucinations\end{tabular} &
  Strict censorship \\ \hline
\end{tabular}%
}
\end{table*}

\section{Potential Opportunities in Information Retrieval with ChatGPT}
In the era of large models, generative models represented by ChatGPT are introducing new perspectives and methodologies for the core task of information retrieval. IR systems aim to extract relevant information from enormous amounts of textual data. Traditional IR systems often rely on keyword matching. However, with the advent of neural networks and deep learning, IR is progressively evolving towards semantic-based retrieval \cite{DBLP:journals/ipm/PanWHHCC22}.

The deep neural networks of GPT-X enable a profound understanding of text semantics, enhancing the precision in semantic-level retrieval beyond traditional keyword-level text matching. Their generative framework allows for the formulation of precise query expressions and the generation of descriptive retrieval results, enhancing the flexibility and expressiveness of IR. With zero or few-shot learning capabilities where models require little to no training data, these models reduce the necessity for extensive annotated data, making complex retrieval tasks more manageable. The end-to-end training methodology minimizes error propagation and directly optimizes performance from input to output, improving retrieval accuracy and efficiency. Furthermore, the potential for multimodal information retrieval extends the scope beyond text to encompass images and videos, offering richer and more accurate retrieval results. Lastly, integrating knowledge graphs leverages structured knowledge in the retrieval process, simultaneously aiding in the construction and updating of knowledge graphs, thus providing a richer knowledge base for IR.

\subsection{Information Extraction}
Information Extraction (IE) is a fundamental task in information retrieval, encompassing sub-tasks such as named entity recognition (NER) and event extraction (EE). IE has evolved significantly over the years. Initially, the focus was on structured and semi-structured data extraction, employing various techniques, tools, and systems to extract useful information automatically \cite{cowie1996information}. Early IE systems were primarily rule-based, relied on a large amount of human involvement, and were tailored for specific domains like chemical or medical search \cite{sarawagi2008information, balke2012introduction, DBLP:journals/sigir/LupuHZT09, DBLP:conf/trec/LupuPHZT09, DBLP:conf/sigir/HuangH09, DBLP:journals/tkde/YinHLZ13}.

Transitioning into the contemporary period, the field has seen a shift towards employing deep learning technologies, which excel at extracting structured information from unstructured text without being confined to a particular domain \cite{adnan2019analytical, rahman2022assessment}. The core idea of deep learning is to extract features from the original data, moving from low-level to high-level and from the concrete to the abstract through a series of non-linear transformations in a data-driven manner. These methods have significantly improved the advanced levels of various fields, including speech recognition, visual object recognition, and object detection, showcasing the efficacy of deep learning in handling complex IE tasks \cite{yang2022survey}.

Moreover, researchers hope these large-scale language models can process text efficiently and extract valuable information without the necessity for retraining, potentially replacing manual annotation. However, multiple extensive IE experiments on ChatGPT show a significant performance gap between ChatGPT and state-of-the-art (SOTA) results on datasets with zero/few-shot IE sub-tasks \cite{wei2023zero, han2023information, yuan2023zero, li2023evaluating}. Although the results are unsatisfactory, they spark new research perspectives in IE, such as the possibility that IE tasks can be decomposed into multiple simpler subtasks \cite{wei2023zero}, a rethinking of the evaluation strategy might reflect a more accurate performance of ChatGPT \cite{han2023information}, and ChatGPT's performance can be significantly improved by prompt engineering \cite{yuan2023zero}.

\subsection{Text Classification}
In exploring text classification tasks in the era of large language models, it's pertinent first to introduce the traditional and prevalent methodologies in text classification. Traditional text classification approaches generally rely on statistical learning paradigms such as Naive Bayes and K-Nearest Neighbors \cite{DBLP:conf/iral/PengHSW03, DBLP:journals/trs/AnHHC04, DBLP:journals/isci/ZhouCHHH20}. These methods entail substantial effort in feature engineering to construct meaningful representations of text. Subsequently, with the advent of deep neural networks, models like RNNs, CNNs, and Graph Neural Networks (GNNs) \cite{yao2019graph} have emerged as mainstream paradigms, significantly automating the construction of rich semantic representations of text.

Entering the era of LLMs, models like ChatGPT have markedly impacted text classification tasks. These models achieve high-quality text semantic modeling from massive text corpora through supervised pre-training techniques, substantially enhancing the performance in text classification tasks. Particularly in addressing open-domain tasks, domain adaptation, few-shot (where models learn from a small set of labeled examples), and zero-shot (where models generalize to unseen classes) problems, these large models exhibit impressive performance and exceptional generalization capabilities \cite{soni2023comparing, chen2023robust, nogueira2019multistage, zamani2020generating}.

ChatGPT can be utilized to undertake a knowledge graph extraction task to obtain refined and structured knowledge from raw data. The collected knowledge is then transformed into a graph, which is subsequently utilized to train an interpretable linear classifier to render predictions, exhibiting impressive performance \cite{shi2023chatgraph}.

In scenarios with few or zero examples, LLMs leverage pre-trained knowledge to achieve satisfactory classification outcomes, mitigating the dependency on large labeled datasets inherent in traditional methods. This capability is invaluable in domains encumbered by limited training data due to costly and labor-intensive annotation processes \cite{zhao2023chatagri}. In addition, high-quality categorization lays a solid foundation for accurate and efficient annotation, thus potentially speeding up the annotation process, reducing costs, and improving the overall quality of the annotated data, which greatly benefits the text annotation task \cite{gilardi2023chatgpt}.

From an information retrieval perspective, text classification serves as a crucial mechanism for ranking and categorizing textual data, aiding in the efficient retrieval and management of information. Combining the knowledge graph and few-shot learning capabilities based on LLMs, text classification tasks can extract and utilize relevant information from extensive data, achieving more accurate and efficient categorization.

\subsection{Document Ranking}
Document ranking is a crucial process in information retrieval systems, determining the order in which retrieved documents are presented based on their estimated relevance to a query. Historically, the methodologies employed for document ranking have predominantly centered on term-based matching, leveraging standard techniques such as Term Frequency-Inverse Document Frequency (TF-IDF) and BM25 \cite{robertson1995okapi}. These traditional approaches assess the significance of terms within documents and their corresponding relevance to the query at hand \cite{DBLP:journals/tois/ZhaoHY14, DBLP:conf/sigir/ZhaoHH11}. However, they often fall short in capturing the semantic relationships between terms and may overlook contextual relevance, which is increasingly important in refining the precision of document retrieval.

Transitioning into the modern era, machine learning has found a foothold in document ranking through methods like Learning to Rank \cite{liu2009learning}, which predicts a relevance score for each document-query pair, ranking documents accordingly. Thereafter, deep learning models started gaining traction. CNNs, RNNs, and attention-based mechanisms such as BERT \cite{DBLP:conf/naacl/DevlinCLT19} have been employed to enhance the representation of text data and improve the understanding of natural language queries \cite{DBLP:conf/lrec/LaskarHH20}. Recently, the focus has also shifted towards dense retrieval and re-ranking models \cite{karpukhin2020dense}. Dense retrieval models propose a more accurate approach to document ranking tasks by embedding both documents and queries in a continuous vector space. Re-rankers take an initial set of retrieved candidates and re-sort them based on relevance scores, ensuring a more reliable list of results in response to a query.

The advent of large language models has opened new possibilities for document ranking in IR. Investigations have revealed that ChatGPT can deliver competitive or even superior ranking performance compared to supervised methods on popular IR benchmarks when properly instructed \cite{wang2023can, sun2023chatgpt}. The emergence of GPT-4 has further pushed the boundaries, showcasing AI-driven document ranking, significantly impacting the search engine domain \cite{sun2023chatgpt}. In addition, a human-involved experiment comparing the search performance and user experience of ChatGPT and Google Search points to practical insights. Although ChatGPT cannot always outperform Google Search, it considerably enhances work efficiency and increases user satisfaction \cite{xu2023chatgpt}.

Furthermore, domain-specific document ranking emerges as a promising area for the application of GPT-4. Presently, ranking methods heavily rely on training data and fine-tuning. However, the scarcity of high-quality annotated datasets in specialized domains such as medicine and law poses a significant challenge, impeding the efficacy of deploying pre-trained models for ranking documents \cite{yzecai2023}. LLMs like GPT-4, endowed with expansive knowledge and pronounced generalization capability due to their vast training data spectrum, present a viable solution. These models hold the potential to serve as data augmentation tools in such contexts, synthesizing pseudo-label data that could improve the performance of retrieval models in data-scarce situations \cite{wang2023query2doc, gao2022precise}. By generating synthetic yet relevant data, GPT-4 could significantly enhance the model's ability to accurately rank documents in domain-specific scenarios, thereby bridging the data gap and facilitating improved performance in document retrieval tasks.

\subsection{Conversational Search}
Conversational Search (CS) has significantly evolved over the years, transitioning from rule-based models to the more advanced machine learning and deep learning models prevalent today \cite{DBLP:journals/csur/KeyvanH23, DBLP:journals/tois/ZouHRK23}. Traditionally, it is divided into two main subtasks: task-oriented and open-dialog/interactive tasks. Task-oriented conversational IR (Information Retrieval) systems employed a pipeline approach, integrating several modules like intent recognition, dialogue management, and response generation to handle user interactions \cite{DBLP:journals/tois/ZhaoHDCX22}. Conversely, open-domain conversational IR systems aim to engage users in more social and less goal-directed conversations. Initially, these systems relied on retrieval-based approaches, but the advent of generative models allowed for more fluid and natural responses \cite{DBLP:conf/sigir/ZouXGZ0HY20, DBLP:conf/sigir/XiaHXZYH22}. They function like an IR system, extracting related information from a pre-designed database.

The introduction of transformer-based models, such as OpenAI's ChatGPT, Anthropic AI's Claude, and Google's LaMDA \cite{thoppilan2022lamda}, marked a paradigm shift in the domain of CS. These models' capability to generate human-like text based on a given context has expanded the horizons of what's possible in task-oriented and open-domain CS systems.

Several opportunities arise as the field advances with contributions from models like ChatGPT and GPT-4. Thanks to these models' impressive intent understanding, semantic parsing, and API integration capabilities, the union of task-oriented and open-domain dialogues under a single technical framework is now attainable. This union could lead to the development of CS systems that are not only functional but also emotionally intelligent, catering to the practical needs of users. Moreover, the pursuit of creating more personalized CS systems remains a significant area of research and development. Advancements in these areas are expected to push CS systems closer to delivering a truly human-like and enriching conversational experience.

\subsection{Multimodal Retrieval}
In the realm of multimodal retrieval, the transition from traditional methods to cutting-edge techniques showcases a remarkable development. Initially, traditional multimodal retrieval predominantly fell under the Nearest Neighbor (NN) problem \cite{hashing8963910}. However, these methods struggled to bridge the semantic gap between low-level features (such as color, texture, and shape) and users' high-level informational needs. As technology advanced, the focus shifted towards crafting unified representations for data across different modalities, such as text, images, audio, and video, aiming to foster seamless and enriched interactions between these modalities \cite{zhao2023retrieving}.

The field then embraced cross-modal retrieval, emphasizing the importance of modeling relationships between different modalities. This approach allowed users to retrieve desired information by submitting data in one modality to fetch related data in another, marking a significant stride towards enhancing accuracy and scalability in retrieval \cite{wang2016comprehensive, 10.1145/3539618.3591687}. Additionally, the emergence of retrieval-augmented multimodal models began integrating external knowledge more scalably and modularly. For a given input text, such models use a retriever to fetch relevant documents from external sources and a generator (often a language model) to produce predictions based on the acquired information. Typically, these external sources include text corpora and structured knowledge bases. However, retrieval-augmented methods were initially researched for text, and extending them to the multimodal domain remains challenging. The main difficulty lies in the design of the retriever and generator that can handle multimodal documents containing both images and text. 

Addressing this challenge, the Retrieval-Augmented Text-to-Image Generator (Re-Imagen) \cite{chen2022reimagen} represents a significant advancement. Utilizing a diffusion-based method, this model generates high-fidelity images that are remarkably accurate, even when depicting entities not previously encountered. The process hinges on the effective use of information retrieved from external sources, enabling the creation of visually precise representations. Similarly, the Multimodal Retrieval-Augmented Transformer (MuRAG) \cite{chen2022murag} focuses on answering natural language questions using image retrieval methods. Although these works concentrate on generating single modalities (text or image), RA-CM3 \cite{DBLP:conf/icml/YasunagaAS0LLLZ23} proposed a comprehensive and unified model capable of retrieving and generating both images and text. Notably, the generator model develops capabilities such as controlled image generation in a contextual learning framework through retrieval-enhanced training. 

The debut of GPT-4 notably impacted the field of multimodal retrieval, ushering in an era closer to human-like AI. GPT-4 is a large multimodal model capable of processing both text and image inputs while delivering text outputs, pushing closer to human-level performance on various benchmarks, albeit with certain limitations in real-world scenarios \cite{li2023prompt}. Conversely, ChatGPT has been empowered by GPT-4V(ision) \cite{openai2023gpt4v}, boosting its multimodal capabilities. For instance, the integration of DALL-E 3\footnote{https://openai.com/dall-e-3} with ChatGPT facilitates smoother interaction, where ChatGPT aids in crafting precise prompts for DALL-E 3, turning user ideas into vibrant AI-generated art.

The arrival of large language models marked a significant milestone in bridging the semantic gap between multiple types of information, paving the way for more intuitive and rich interactions across diverse data modalities. In recommendation systems, LLMs have shown immense promise \cite{wang2023generative}. They foster a more comprehensive understanding of user preferences and behaviors by integrating information from various sources and modalities. For example, a recommendation system powered by a multimodal LLM can analyze textual reviews, image-based preferences, and purchase histories to generate more accurate and personalized product recommendations. Moreover, by understanding the semantic relationships between different items and user interactions, these models can provide a more enriched and personalized user experience, thereby enhancing user satisfaction and engagement.

Similarly, the medical field has seen substantial advancements by incorporating LLMs \cite{li2023chatgpt, huang2023chatgpt}. In clinical settings, multimodal LLMs can assist in synthesizing information from diverse sources such as electronic health records, medical imaging, and genomic data to provide more comprehensive and personalized insights. This holds vast potential to support diagnostic processes, treatment planning, and personalized medicine. For instance, integrating textual clinical notes with medical imaging data can empower clinicians with a more holistic understanding of a patient's condition, enabling better-informed decision-making.

\section{Unresolved Challenges in Information Retrieval with ChatGPT}
ChatGPT proves the duality of technological advances in AI. On the one side, it can greatly enhance the productivity of users from all walks of life thanks to its excellent language comprehension and generation capabilities. Whether in education, business, or personal assistance, ChatGPT is a powerful tool that facilitates task completion, inspires creativity, and helps make correct decisions.

On the flip side, it reveals the ethical dilemmas associated with misinformation, disinformation, and the potential misuse possibilities of fabricating deceptive or harmful content. Its remarkable ability to produce realistic text blurs the boundaries of information authenticity, making it challenging for individuals to discern between real and fake content. These potential risks highlight the limitless possibilities of ChatGPT while also emphasizing the need to navigate the ethical regulation that accompanies such groundbreaking innovations.

\subsection{Hallucination}
The challenge of hallucination in large language models, underscored by Google AI researchers in 2018 \cite{lee2018hallucinations}, presents a formidable hurdle to their deployment. Hallucination, a phenomenon where models generate convincing yet factually incorrect or misleading content, harbors serious risks. This is particularly concerning in critical applications such as decision-making, where the propagation of false information can lead to adverse outcomes \cite{bang2023multitask, sallam2023chatgpt}. OpenAI, the developer of ChatGPT, has recognized the concerns regarding the model's propensity for factual inaccuracies and is actively pursuing measures to mitigate this issue\footnote{https://www.technologyreview.com/2023/03/03/1069311/inside-story-oral-history-how-chatgpt-built-openai}.

Information retrieval strategies are poised to be instrumental in addressing the hallucination challenge. A viable approach could be establishing a continuous feedback loop wherein the model's outputs are rigorously evaluated, and refinements are made based on identified inaccuracies. This iterative process aims to bolster the model's accuracy and reliability over time. Specifically, integrating IR models to work in tandem with LLMs could present a robust solution \cite{gao2022rarr}. By augmenting LLMs with updated and accurate information extracted from external sources, IR models can potentially curtail the generation of factually inaccurate responses, thus mitigating the occurrence of hallucinations. 

\subsection{Ethical Issues and Safety}
The ethical and safety concerns surrounding ChatGPT are multi-faceted, arising from their profound language understanding and generation capabilities. As advanced iterations of language models developed by OpenAI, these models harbor significant expectations alongside concerns due to their potential transformative impact on society \cite{zhuo2023red}.

The expansive training data and the complex nature of these models introduce risks associated with bias and fairness. The training material, sourced from human-generated content, may inadvertently perpetuate existing societal biases. Instances where models exhibited gender or racial biases are emblematic of this problem. These biases can manifest across various applications, potentially leading to unfair or discriminatory outcomes \cite{dash2023chatgpt}.

Moreover, the emergence of generative AI poses challenges related to misinformation and abuse. Their ability to generate text can be leveraged to fabricate misleading information, contribute to online misinformation campaigns, or even generate harmful or abusive content. The lack of source attribution in responses generated by ChatGPT exacerbates this issue, as users may struggle to discern the veracity of the generated content \cite{ray2023chatgpt}. The potential misuse of generative AI for criminal activities such as fraud or harassment is another significant concern. LLMs can be employed to create realistic fake content for nefarious purposes, thereby reducing costs and increasing the efficiency of executing fraudulent activities.

In an IR system, while relevance is often prioritized, this can lead to insufficient diversity in the results \cite{DBLP:conf/sigir/HuangH09, DBLP:journals/tkde/YinHLZ13}. Frequently, the most prominent subtopic groups dominate the search results, marginalizing minority topics. This imbalance can cause users to exert extra effort to find items related to less common topics, leading to a partial and skewed information result. Moreover, the tendency of users to click primarily on top search results can facilitate a cycle of unfairness. Ranking or recommendation algorithms incorporating user feedback tend to maintain these items' top positions, creating a positive feedback loop for unfairness issues. This is particularly problematic in systems like ChatGPT, where an initial response with an ethical issue can be challenging to correct internally. If users trust such a response, repeated interactions can exacerbate the issue, contributing to the development of bias. Quantifying bias in terms of gender and age can be beneficial to address these challenges \cite{10.1145/3292500.3330691}. A fairness-aware ranking algorithm that accounts for these factors can have a positive impact. Additionally, considering fairness as an optimization problem opens up new approaches \cite{10.1145/3341981.3344215}. Implementing a fairness-constrained reinforcement learning algorithm can help balance relevance with the need for diversity and fairness in IR systems \cite{10.1145/3437963.3441824}.

Furthermore, the scalability of these models amplifies privacy concerns. As models become larger and necessitate more computational resources, the need to offload processing to cloud servers escalates. This centralization can heighten the risk of data breaches and misuse of personal information, especially if adequate measures are not in place to secure user data.

Overall, the array of ethical and safety concerns emanating from the deployment of ChatGPT underscores the imperative of diligent oversight, robust regulatory frameworks, and continuous dialogue among stakeholders to ensure the responsible development and use of these transformative technologies.

\subsection{Interpretability}
As language models become more complex with increased parameters and depth, their decision-making processes become less interpretable. This complexity also challenges understanding the vector and parameter representations within deep neural networks \cite{DBLP:conf/ijcnlp/DanilevskyQAKKS20}.

Characterizing large language models as ``black-box'' models summarizes this fundamental challenge in deploying and trusting these systems \cite{guidotti2018survey}. While the user can observe the inputs and outputs, the intricacies of the processes in between remain hidden, preventing a clear understanding of how the model derives a particular output from a given input. This opacity extends to an inability to discern what aspects of the input data the model considers important, obscuring interpretability.

The main reason for this challenge is that while LLMs are good at recognizing patterns and correlations in data, they lack a grasp of causality \cite{xu2019explainable}. This inadequacy is particularly evident in decision-making. Moreover, LLMs are prone to inherit biases present in the training data, which highlights another dimension of the interpretability challenge. Any bias may permeate the model's behavior, leading to anomalous or unfair results. Diagnosing and mitigating these biases becomes difficult without a clear window into the model's inner workings.

The challenge of interpretability is further exacerbated by the unpredictability of LLMs in the face of new or adversarial inputs. These models may exhibit erratic behavior in the face of unexpected input scenarios, which is difficult to address without an interpretability framework. Improving the interpretability of LLMs is, therefore, not just an academic exploration but a pragmatic need to ensure responsible and credible deployment of these models, especially as they enter increasingly sensitive and critical domains. Uncovering the ``black box'' nature of LLM and building robust interpretability frameworks is necessary for developing machine learning and AI. 

Retrieval-Enhanced Machine Learning \cite{10.1145/3477495.3531722} presents a promising approach to addressing the issue of interpretability. In pre-trained language models, the training knowledge is embedded within the learned model parameters, making it difficult to understand model predictions. In contrast, when the reasoning process relies on retrieved information, predictions can be directly linked to specific data, typically stored in an accessible text format. This feature improves the interpretability of the model's outputs. Additionally, Aspect Learning \cite{10.1145/3534678.3539137} can further enhance interpretability. By incorporating aspects, the model not only grasps general language semantics, like other pre-trained models, but also acquires domain-specific knowledge, enabling it to identify aspects relevant to a particular domain. These ``explicit aspects'' significantly improve interpretability, as the retrieved documents are expected to share similar aspects (or categories) with the input query.

OpenAI has initiated efforts to automate the interpretability of large language models by using GPT-4 itself to generate and score explanations of neuron behavior in other language models \cite{bills2023language}. This initiative aims to uncover how different parts of the neural network operate, although the technique still struggles with larger models, indicating room for improvement. The initiative by OpenAI represents a significant stride towards demystifying the operations of LLMs, hoping to foster more responsible and effective use of these powerful tools in various domains.

\section{Conclusions and Future Directions}
ChatGPT signifies a remarkable stride in Generative AI, enriching multiple information retrieval tasks. They excel in understanding and generating textual content, with applications extending to various practical and academic domains such as healthcare, education, and programming, thus reshaping traditional paradigms. However, this advancement isn't without challenges. 

Ethical dilemmas such as misinformation, disinformation, and potential misuse for harmful content generation pose serious concerns. The issue of hallucination, generating incorrect or misleading content, highlights the need for robust mechanisms to ensure accuracy and reliability. Furthermore, the challenge of interpretability remains a substantial hurdle. The ``black box'' nature of these models hinders transparency in their decision-making processes, which is essential for responsible AI deployment, especially in critical domains. 

In addressing these challenges,  recent works in IR have made strides in these areas. We note that fairness retrieval methods have shown the potential to mitigate biases in PLLMs, promoting more equitable and unbiased content generation. Additionally, the application of retrieval-enhanced learning methods has been identified as beneficial in tackling interpretability issues. By integrating context-rich information into the learning process, these methods can provide insights into the decision-making mechanisms of these complex models.

The advent of ChatGPT embodies the broader narrative of AI development, filled with promises of technological innovation and the imperative of addressing ethical, safety, and privacy challenges. Continued research and proactive steps to mitigate these challenges while exploring new ways to harness the power of these models responsibly will help navigate the complexities of AI. Collaborative efforts among researchers, practitioners, and policymakers are pivotal in realizing a future where AI significantly enhances human capabilities while preserving ethical and social values.

\begin{acks}
We express our sincere gratitude to the reviewers for their insightful comments and to the editor for their valuable assistance, both of which have significantly contributed to the enhancement of this paper. This research is supported by the Natural Sciences and Engineering Research Council (NSERC) of Canada and the York Research Chairs (YRC) program.
\end{acks}

\nocite{*}
\bibliographystyle{ios1}           
\bibliography{ir_survey.bib}        

\end{document}